# Efficient Subsidy Targeting in the Health Insurance Marketplaces

**Authors:** Coleman Drake, PhD,[1] Mark Meiselbach, PhD,[2] Daniel Polsky, PhD[2]

[1] University of Pittsburgh School of Public Health

[2] Johns Hopkins University Bloomberg School of Public Health

**Corresponding Author**: Coleman Drake, PhD. 130 De Soto Street, Pittsburgh, PA15261. Email: cdrake@pitt.edu.

**Date:** October 27, 2025

**Abstract Word Count:** 227/250

**Word Count**: 5,989/6,000

**Acknowledgments**: Drake and Meiselbach acknowledge support for this work from the Commonwealth Fund. Drake and Polsky acknowledge support for this work from the Agency for Health Care Research and Quality (AHRQ) under award R01HS030083. We are grateful to the Maryland Health Benefit Exchange (MHBE) for providing the data for our analysis, and to Amelia Marcus, Johanna Fabian-Marks, and Michele Eberle for their assistance in obtaining the data. The opinions expressed in this manuscript do not necessarily reflect those of AHRQ, the Commonwealth Fund, or MHBE. We also are grateful to Dylan Nagy and David Anderson for providing programming support and technical expertise, respectively. Additionally, we are thankful for helpful comments and discussion from Graeme Peterson, Mark Shepard, Daniel Sacks, and Travis Donahoe.

# Abstract

Enrollment in the Health Insurance Marketplaces created by the Affordable Care Act reached an all-time high of approximately 25 million Americans in 2025, roughly doubling since enhanced premium tax credit subsidies were made available in 2021. The scheduled expiration of enhanced subsidies in 2026 is estimated to leave over seven million Americans without health insurance coverage. Ten states have created supplemental Marketplace subsidies, yet little attention has been paid to how to best structure these subsidies to maximize coverage. Using administrative enrollment data from Maryland's Marketplace, we estimate demand for Marketplace coverage. Then, using estimated parameters and varying budget constraints, we simulate how to optimally allocate supplemental state premium subsidies to mitigate coverage losses from enhanced premium subsidy expiration. We find that premium sensitivity is greatest among enrollees with incomes below 200% of the federal poverty level, where the marginal effect of an additional ten dollars in monthly subsidies on the probability of coverage is approximately 6.5 percentage points, and decreases to roughly 2.5 percentage points above 200% FPL. Simulation results indicate that each $10 million in annual state subsidies could retain roughly 5,000 enrollees, though the cost-effectiveness of these subsidies falls considerably once all enrollees below 200% of the federal poverty level are fully subsidized. We conclude that states are well positioned to mitigate, but not stop, coverage losses from expanded premium tax credit subsidy expiration.

# 1. Introduction

Enrollment in the Health Insurance Marketplaces created by the Affordable Care Act (ACA) reached an all-time high of 25 million people in 2025 eleven years after their creation. Increases in the generosity of Marketplace premium tax credits (PTCs), implemented under the American Rescue Plan Act of (ARPA) and subsequently by the Inflation Reduction Act (IRA), caused Marketplace enrollment to double since 2021. Yet the expanded premium tax credits (ePTCs) in place under the ARPA and the IRA since 2021 are due to expire on January 1$^{st}$, 2026, unless they are extended by Congress. Some 3.9 million Marketplace enrollees are projected to become uninsured in 2026, and 7.4 million by 2030 (Congressional Budget Office 2024), likely leading to decreased access to care and health (Courtemanche et al. 2018a; 2018b; 2020) and increased mortality (Goldin et al. 2020).

In this context, it is important to understand how states can cost-effectively structure supplemental Marketplace premium subsidies to maximize coverage gains. While ten states have created various forms of subsidies, no study has examined how to optimally structure state Marketplace subsidies. Given state policymakers' interest in leveraging their limited budgets to mitigate Marketplace coverage losses (Swindle and Giovanelli 2024; Maryland Health Benefit Exchange 2025), we simulate how to cost-effectively maximize Marketplace coverage as ePTCs expire.

We derive parameters for our simulation by estimating demand for Marketplace coverage using administrative 2022-2024 data from Maryland's Health Insurance Marketplace, the Maryland Health Benefit Exchange (MHBE), alongside data on potential enrollees from the American Community Survey (ACS). We estimate demand using an instrumental variables approach with a simulated instrument (Currie and Gruber 1996)—what subsidized premiums would have been if the ARPA had never created ePTCs. We then recover marginal effects of premium increases on coverage and develop an algorithm that allocates state supplemental subsidies to the most premium sensitive enrollees under annual budgets ranging from $10 to $150 million. Across these budgets, we report coverage gains, average and marginal costs, and how subsidies are distributed across enrollees' income distribution.

Broadly, we find that supplemental subsidies for lower income Marketplace enrollees are relatively cost effective, though retaining coverage for higher income Marketplace enrollees becomes prohibitively costly. For potential enrollees below 200% FPL, the marginal effect of ten dollars in subsidies spent is a roughly 6.5 percentage point increase in the probability of coverage. This marginal effect decreases to about 2.5 percentage points for potential enrollees above 200% FPL, making it significantly costlier, and less cost effective, for the state to increase coverage among higher income potential enrollees. For enrollees below 200% FPL, however, each $10 million in additional state subsidies can retain roughly 5,000 enrollees. Based on what states like Maryland are already spending on supplemental subsidies (Lane 2025; Swindle and Giovanelli 2024), they are well positioned to preserve slightly under half of the coverage gains from the ARPA and the IRA for enrollees with incomes below 200% FPL. Although some states' budgets are more limited, the cost effectiveness of state supplemental subsidies is highest at zero, meaning that even small supplemental schemes could be effective.

Our work contributes to the literature on the design of government-subsidized private health insurance markets in several ways. First, while several prior studies examine how federal policymakers should structure the price-linked subsidies characteristic of these markets (Jaffe and Shepard 2020; Finkelstein et al. 2019; Tebaldi 2025), ours is the first, to our knowledge, to study how states can supplement and respond to price-linked subsidies. Our findings provide timely and policy relevant information to state policymakers considering altering and creating state supplemental subsidies in response to ePTC expiration. This type of federal-state interaction inherent in these supplemental subsidies is also relevant to other countries with both a federal system of government and government-subsidized private health insurance markets, notably Australia (Liu and Zhang 2023), Belgium (Schokkaert et al. 2010), Germany (Bünnings et al. 2019), and Switzerland (Ginneken et al. 2017).

Second, our study is the first post-ARPA, post-pandemic study to estimate extensive margin elasticities and semi-elasticities of coverage. Here, we find demand for Marketplace coverage is considerably more premium sensitive than in prior, pre-ARPA/IRA Marketplace studies (Goldin et al. 2020; Saltzman 2019; Panhans 2019; Tebaldi 2025; Drake and Anderson 2020; Tebaldi et al. 2023; Soni 2022; Frean et al. 2017; Hopkins et al. 2025; Drake et al. 2023) and closer to those of the pre-ACA Massachusetts Marketplace (Finkelstein et al. 2019; Shepard 2022; Hackmann et al. 2015; 2012). These differences may simply reflect that premium sensitivity is higher in markets with more lower income enrollees. The ARPA/IRA ePTCs led to decreases in the income of the mean Marketplace enrollee by increasing lower income coverage. This made the federal Marketplaces income distribution more like the Massachusetts pre-ACA Marketplace, which had relatively lower income enrollees due to its lower income cap on subsidies.

Our paper proceeds as follows. Section 2 describes the regulatory structure of the Marketplaces, focusing on premium subsidies. Section 3 describes the data. Section 4 develops and estimates a model of demand for coverage and reports results. Section 5 simulates coverage retention and costs when states optimally allocate funds to preserve a portion of ePTC funding and discusses policy implications. Section 6 concludes.

# 2. Background

**2.1. The Health Insurance Marketplaces**

The Health Insurance Marketplaces, created by the *Affordable Care Act (ACA)* in 2010 and implemented in 2014, provide partially community-rated health plans with guaranteed issue and minimum benefits. People without affordable offers of insurance from their employers or eligibility for public insurance programs, namely Medicaid and Medicare, receive premium tax credit subsidies to purchase Marketplace coverage from the federal government. Several policy changes at the federal and state level have modified premium tax credit size and eligibility since 2014. For example, in mid-2021 the *American Rescue Plan Act (ARPA)* and subsequent *Inflation Reduction Act (IRA)* expanded subsidies through 2025. Ten states, including Maryland, have supplemented the federal subsidies using state funds (Swindle and Giovanelli 2024).

Each state has its own Marketplace. Historically, most states used the federal platform, healthcare.gov, rather than administering their own Marketplace. However, the number of states using their own state-based Marketplaces has increased to 20 as of 2025; two more will by 2027 (CMS 2024). Maryland, which we examine in this study, has had a state-based Marketplace, the *Maryland Health Benefit Exchange (MHBE)*, since 2014.

**2.2. Premiums and Benefits**

Marketplace insurers set their plans' premiums in coordination with states or the federal government based on experience rating and generosity of coverage. Insurers can vary premiums across clusters of counties known as geographic rating areas. Insurers set one premium for each plan offered in each rating area. Those premiums are multiplied by an age adjustment factor based on the age of each household member enrolling in coverage. Age adjustment varies on a three-to-one basis, where age 21 is set at one and age 64 and above are set at three. Maryland and nearly all states use the default age adjustment schedule set by the federal government.[1]

Each Marketplace plan must have a *metal level* roughly corresponding to an actuarial value. Bronze plans have about 60% actuarial value, silver have 70%, and gold 80%.[2] Enrollees with incomes at or below 250% FPL receive *cost-sharing reduction (CSR)* subsidies that increase the actuarial value of silver plans to 94% for enrollees with incomes between 100-150% FPL, to 87% for those between 150-200% FPL, and 73% from 200-250% FPL. The 94% and 87% subsidies bifurcate the Marketplaces such that enrollees with incomes below 200% FPL overwhelmingly enroll in silver plans, while those above 200% FPL enroll in a mixture of bronze, silver, and gold plans. The 73% CSR subsidies have little impact (DeLeire et al. 2017).

**2.3. Federal Marketplace Premium Tax Credits**

Federal advanced *premium tax credits (PTCs)* are price-linked subsidies (Jaffe and Shepard 2020) that limit the amount eligible Marketplace enrollees must pay for Marketplace coverage as a percentage of their income; PTCs cover the residual. Premium tax credits are indexed to the premium of the *benchmark plan*—the second lowest premium silver plan available in a county-year. By covering residual premiums, PTCs set the benchmark plans' premium equal to an *expected contribution percentage*, a sliding scale percentage of an enrollee's *modified adjusted gross income (MAGI)*. Although PTCs are calculated based on the benchmark plan's premium, they can be applied towards the premium of any other plan. If enrollees select a plan with premium below the benchmark plan's, they can pay less for coverage than their expected contribution percentage. PTCs cannot reduce premiums below zero dollars.

---

[1] Enrollees indicating that they smoke when applying for coverage are charged an additional 50% for coverage that is not reimbursed by subsidies. Practically, there is little verification of smoking status and few enrollees indicate they smoke.
[2] Platinum plans with 90% actuarial exist, though insurers frequently do not offer them and enrollment is minimal. Catastrophic plans also exist with actuarial value slightly below 60%, but they are ineligible for premium subsidies and, as such, enrollment is very low among subsidy-eligible enrollees.

Expected contribution percentages ranged from roughly two to ten percent of MAGI from 2014 through early 2021, with eligibility for PTCs expiring at 400% FPL. The ARPA and the IRA's *enhanced premium tax credits (ePTCs)* worked by reducing expected contribution percentages from zero percent to 8.5% and removing the 400% FPL eligibility cap beginning from 2021 through 2025. **Appendix Figure A1** illustrates how the ARPA/IRA changed expected contribution percentages and expected contribution amounts—what an enrollee with PTCs will pay for the benchmark plan after applying PTCs—across the income distribution. Subsidies will reduce to pre-ARPA levels when ePTCs expire in 2026.

## 2.4. State Subsidy Programs

Ten states currently supplement federal subsidies with state-funded subsidies (Swindle and Giovanelli 2024), as shown in **Appendix Table A1**. These subsidies typically increase the generosity of federal premium subsidies by either reducing expected contribution percentages, such as Maryland and Vermont, or providing fixed dollar amount subsidies, such as New Jersey and Washington. State supplemental subsidies are typically conditioned on eligibility for federal PTCs, though some states have expanded PTC eligibility in cases where people are not eligible. California and Minnesota, for example, provided PTCs to enrollees above 400% FPL before the ARPA made them available, and Colorado provides subsidies to immigrants that would not otherwise be eligible. Supplemental subsidy generosity is typically based on income like federal PTCs, although Maryland conditioned supplemental subsidy on age.

Maryland, which we examine in this study, began supplementing federal PTCs in 2022 with the creation of its *Young Adult Subsidy (YAS)* program. As it was implemented in 2022, the YAS reduced the expected contribution percentages of enrollees aged 18 to 30 by 2.5 percentage points. It phased out per additional year of age by 0.5 percentage points, ending at age 35. In 2023, the YAS was expanded such that the phase out began at age 34 and ended at age 38.[3] The YAS is financed through Maryland's 1332 waiver state reinsurance program, which is authorized to operate through 2028 (CMS 2025).

**2.5. Future Policy Changes and Federal-State Subsidy Interaction**

As of September 2025, it appears unlikely Congress will extend ARPA/IRA enhanced premium tax credits. States with supplemental Marketplace subsidy programs thus find themselves in a situation where the costs, benefits, and efficiency of their supplemental subsidies will change as federal subsidies revert to pre-ARPA levels in 2026. We consider such subsidy changes in our analyses below.

The costs of providing federal and state supplemental subsidies are endogenous—they will change in response each other's generosity levels. Consider a state that increases federal subsidy spending by 20%. Consumers whose enrollment is marginal to the existence of the state subsidies will also receive federal PTCs, increasing the cost of providing federal PTCs by the level of federal PTCs times the number of new enrollees. The same may occur in reverse. For example, ePTC elimination will likely reduce states' supplemental subsidy costs by reducing demand for Marketplace coverage.

# 3. Data

**3.1. Data Sources**

---

[3] The YAS also was modified in 2023 so that it could eliminate the portion of plans' premiums due to non-Essential Health Benefits; federal PTCs cannot do this. Maryland mandates that Marketplace plans cover non-Hyde abortion benefits, so prior to 2023 this mandate meant that zero-premium MHBE plans did not exist. Allowing the YAS to cover these benefits, alongside other non-Essential Health Benefits like adult and vision coverage, meant that zero-premium plans became available in Maryland for the first time.

Our primary data are 2022-2024 administrative enrollment data and plan offering data provided by Maryland's state-based Marketplace, the Maryland Health Benefit Exchange (MHBE). The enrollment data list all enrollees' plan selections, federal and state premium subsidies awarded and received, and coverage start and end dates. They also provide demographics, including age, income as a percentage of the federal poverty level (FPL), anonymized individual and household identifiers, and county of residence. Plan data identify standard plan characteristics, including premiums, metal levels, and counties and rating areas in which they are offered. We use the plan data to identify the benchmark plans and their premiums in each county-year. We use additional data sources for subsidy calculation as reported in Section 3.3.

We supplement the MHBE data with 2022-2023 American Community Survey data (Ruggles et al. 2025), which we use to identify subsidy-eligible potential Marketplace enrollees. The ACS surveys 1% of the U.S. population each year. These data include the same demographic characteristics as the MHBE data. We convert reported income in the ACS to income as a percentage of the FPL as defined by ASPE (Assistant Secretary for Planning and Evaluation 2024). We convert public use microdata areas, the geography reported in the ACS, to counties using Geocorr (Missouri Census Data Center 2022).

### 3.2. Sample Selection

We limit our sample to premium subsidy-eligible enrollees and potential enrollees in the traditional ACA premium subsidy eligibility range, 138-400% FPL. This will be the only population eligible for PTCs when IRA-funded ePTCs expire in 2026. We also exclude persons aged 65 and above, who are typically covered by Medicare, and American Indian and Alaskan Natives subsidies, who receive different types of PTCs than other Marketplace enrollees. We exclude less than one percent of MHBE data where premium subsidy data are missing.

Following Saltzman (2019), we construct our ACS sample by applying sample restrictions described above and limiting the ACS sample to uninsured persons. We use ACS survey weights to impute the size of the uninsured, subsidy-eligible population.

**3.3. Measures**

Our primary outcome is enrollment in MHBE, and our "treatment" is post-subsidy premiums. We measure the affordability of subsidized Marketplace coverage as the post-PTC premium of the lowest premium silver plan available in a market (i.e., a rating area-year). We use three data sources to calculate this measure: (1) Marketplace coverage expected contribution percentages by FPL and year from the IRS (Internal Revenue Service 2025); (2) Marketplace age adjustment factors reported by CMS (Centers for Medicare & Medicaid Services 2024); and (3) the MHBE plan offering data.

First, we use the MHBE plan data to identify the lowest and second-lowest premium benchmark silver plan in each rating area-year. Second, we apply age adjustment factors to determine these plans' pre-subsidy premiums for enrollees and potential enrollees of different ages. Third, we identify the post-subsidy premium of the benchmark plan as enrollees' and potential enrollees' expected contribution amounts, and their federal subsidies as the difference between pre-subsidy premiums and expected contribution amounts. Fourth, we calculate the post-subsidy premiums of the lowest silver plan as premiums less subsidies. Post-subsidy premiums have a floor of zero. Our calculations also incorporate Maryland's Young Adult subsidy program.

Bronze plans are typically cheaper than silver plans and thus represent the absolute minimum point at which marginal enrollees can obtain coverage; however, we prefer using minimum silver premiums for several reasons. First, silver coverage is the standard level of Marketplace coverage in that subsidies are based on the benchmark silver plan. Second, lower income enrollees (i.e., <200% FPL) overwhelmingly select silver plans because cost-sharing reduction subsidies make them far more generous than bronze plans. Third, bronze coverage is essentially catastrophic coverage—bronze and catastrophic plans have roughly the same actuarial values—and thus does not provide the same financial protection and medical access as silver or gold coverage.

**3.4. Sample Characteristics**

**Table 1** provides descriptive statistics on the portion of our sample covered by MHBE. Subsidized MHBE enrollment in the traditionally subsidy-eligible income range (138-400% FPL) increased from approximately 110,900 enrollees in 2022 to 143,100 enrollees in 2024. These increases are consistent with national increases in Marketplace enrollment after ARPA's passage in 2021 (Kaiser Family Foundation 2024). As ePTCs were implemented and enrollment grew, the percentage of relatively younger enrollees increased from 39.5% in 2022 to 44.7% in 2024, suggesting marginal enrollees are relatively healthier. Relatively lower income enrollment (138-200% FPL) also increased from 41.9% to 44.6% over the same time period. Mean pre- and post-PTC premiums and the metal levels of chosen plans exhibited little variation during the study period.

     **Appendix Table A2** compares those insured by MHBE in our sample to uninsured persons eligible for MHBE subsidies identified in the ACS. The size of the two populations is roughly the same by 2024, about 147,000 persons. Uninsured potential enrollees are more likely to be younger, in the 18-39 age group (53.4% vs. 42.2%), and higher income, in the 201-400% FPL group (65.9% vs. 57.3%). As a result of these demographic differences and the PTC subsidy

formula, the uninsured potential enrollee is eligible for a PTC that is, on average, 100 dollars smaller than subsidized MHBE enrollees.

## 4. Coverage Demand Estimation

Our empirical strategy is to estimate demand for MHBE coverage using an instrumental variable (IV) approach. We instrument the post-subsidy premium of the lowest silver plan with what that plan's premium would have been with original ACA PTCs without ePTCs from the ARPA and the IRA.

### 4.1. Instrumental Variables Approach

Our unit of analysis is the person-year. For potential enrollee $i$ in rating area $r$ in year $t$, we estimate the probability of coverage, $\Pr(Cvg_{irt})$, using two-stage least squares with two instruments such that

$$P_{irt}^{ARPA} = \beta_{11} P_{irt}^{ACA} + \beta_{12} P_{irt}^{ACA} FPL_i + \beta_{13} P_{irt}^{ACA} FPL_i^{>200} + \gamma_1 X_{it} + \Theta_{rt} + \mu_{1irt} \tag{1A}$$

$$P_{irt}^{ARPA} FPL_i = \beta_{21} P_{irt}^{ACA} + \beta_{22} P_{irt}^{ACA} FPL_i + \beta_{23} P_{irt}^{ACA} FPL_i^{>200} + \gamma_2 X_{it} + \Theta_{rt} + \mu_{2irt} \tag{1B}$$

$$P_{irt}^{ARPA} FPL_i^{>200} = \beta_{31} P_{irt}^{ACA} + \beta_{32} P_{irt}^{ACA} FPL_i + \beta_{33} P_{irt}^{ACA} FPL_i^{>200} + \gamma_3 X_{it} + \Theta_{rt} + \mu_{3irt} \tag{1C}$$

$$\Pr(Cvg_{irt}) = -\alpha_1 \widehat{P_{irt}^{ARPA}} - \alpha_2 \widehat{P_{irt}^{ARPA} FPL_i} - \alpha_3 \widehat{P_{irt}^{ARPA} FPL_i^{>200}} + \gamma_4 X_{it} + \Theta_{rt} + \omega_{irt}. \tag{1D}$$

We instrument post-ePTC minimum silver premiums potential enrollees were awarded under the ARPA and the IRA during our sample period, $P_{irt}^{ARPA}$, and its interactions with a continuous income term, $FPL_i$, and an indicator for income above 200% FPL, $FPL_i^{>200}$, with simulated post-PTC subsidy minimum silver premiums as they were originally formulated under the ACA prior to ARPA, $P_{irt}^{ACA}$, and equivalent income interactions. We include the $FPL_i^{>200}$ interaction because of the significant differences between Marketplace plan choices at the 200% FPL income threshold due to the presence of CSR subsidies discussed in Section 2 (DeLeire et al. 2017). Our covariates and their functional form are highly similar to Saltzman (2019).

These models also include a vector of demographic controls, $X_{it}$, including continuous FPL and age terms and a female indicator variable, as well as rating area-by-year fixed effects, $\Theta_{rt}$. We cluster the error terms, $\mu_{irt}$ and $\omega_{irt}$, at the household level to address intra-household correlation in health coverage choices.[4] Each MHBE enrollee receives a weight of one; each uninsured ACS respondent is weighted per their survey weight.

### 4.2. Marginal Effects and Elasticities

We convert our estimates to marginal effects and extensive margin premium elasticities and semi-elasticities of coverage. We obtain marginal effects of increasing premiums by one dollar on the probability of coverage by simply multiplying alpha coefficients by 100, using linear combinations of alpha coefficients with the interaction terms. The extensive margin premium elasticity of coverage measures the percentage change enrollment affected by a one percent change in post-subsidy premiums; the semi-elasticity captures the percentage change in enrollment affected by a 100-dollar increase in post-subsidy premiums. We calculate them as

$$\epsilon_F = -\bar{P}_F^{ARPA}(\hat{\alpha}_1 + \hat{\alpha}_2 FPL_F + \hat{\alpha}_3 FPL_F^{\geq 200})/\overline{Cvg}_F \qquad (2A)$$

$$\eta_F = -100(\hat{\alpha}_1 + \hat{\alpha}_2 FPL_F + \hat{\alpha}_3 FPL_F^{\geq 200})/\overline{Cvg}_F, \qquad (2B)$$

where $F$ refers to a specified income range (e.g., 138-150% FPL, 151-200% FPL, whole sample), and $\bar{P}_F^{ARPA}$ and $\overline{Cvg}_F$ are within-income group mean premiums and enrollment rates. We calculate standard errors using the delta method.

### 4.3. Identification

We employ an instrumental variables approach to address several standard concerns regarding premium endogeneity in demand for health insurance. Concerns include adverse selection and other factors affecting coverage decisions, related insurer pricing responses such as within-rating area-year variation in advertising, and supply of and demand for other forms of health insurance (Hackmann et al. 2015; Aizawa and Kim 2018).

---

[4] In the MHBE data, the household is the nuclear family that collectively jointly purchases health insurance. In the ACS data, households expand beyond the nuclear family that jointly purchases health insurance. Accordingly, we use "health insurance units," which capture the same concept as in the MHBE data.

We instrument the post-subsidy minimum silver premiums enrollees actually experienced, $P_{irt}^{ARPA}$ with the post-subsidy minimum silver premiums enrollees would have experienced if the ARPA and IRA had never created expanded subsidies and original ACA subsidies had remained in place, $P_{irt}^{ACA}$. This estimator compares coverage demand across potential enrollees facing varying post-subsidy premiums, using counterfactual post-subsidy premiums that would have existed without the ARPA and IRA as an instrument for observed premiums. This approach is conceptually similar to prior studies that have used counterfactual price instruments (Chetty et al. 2011) and simulated instruments (Cutler and Gruber 1996; Gruber and Simon 2008).

We have two identifying assumptions. First is the instrument condition—that our instruments ACA post-subsidy premiums strongly predict ARPA/IRA post-subsidy premiums. We show that this is the case below in the results. The second identifying assumption is the exclusion restriction—that ACA post-subsidy premiums have no independent influence on enrollment other than through their mechanical relationship to ARPA/IRA post-subsidy premiums. This is highly plausible, primarily because potential enrollees are not presented with any information regarding the instrument when they decide whether to enroll. Unlike the actual post-subsidy premiums that are prominently displayed on the MHBE website, hypothetical premiums that would have existed under prior ACA regulations are not displayed. Plan offerings also were likely not affected by premiums under the old subsidy regime. Even for the 2022 plan year, the ARPA was passed in March 2021, providing insurers with sufficient time to adjust their 2022 plan offerings to the new subsidies. Our use of rating area-year fixed effects increases the robustness of our approach by controlling for unobserved insurance market-by-year-varying phenomena such as advertising and regional health shocks.

### 4.4. Interpretation and Limitations

We interpret our IV estimates as measuring marginal enrollees' price sensitivity to MHBE coverage. These local average treatment effects (LATE) are broad in two senses. First, our IVs are not limited to a subset of our sample because counterfactual ACA post-subsidy premiums exhibit variation across our sample. Second, the premium variation we observe in our data is, by design, equal in magnitude to the state supplemental subsidies we simulate below, meaning that we do not have to make out of sample predictions to simulate responses to different subsidy regimes. The standard limitation applies that our LATE estimates reflect the behavior of enrollees who were moved to enroll due to price changes, though it is these marginal enrollees whose behavior we seek to predict.

      We cannot fully account for household composition in our approach because we often do not observe children in the MHBE data due to their higher Medicaid eligibility through Maryland's version of the Children's Health Program (CHIP), the Maryland Children's Health Program. This scenario is fairly common, as the income limit for the Maryland CHIP for children under age 19 is roughly 211% FPL, and low premium plans are available at higher incomes (Maryland Department of Health 2025). Our use of an IV approach helps to reduce this source of measurement bias. This limitation is inherent in all Marketplace studies using administrative data not also linked to administrative Medicaid data.

### 4.5. Results

We report OLS and IV estimates in **Table 2**. As expected, the IV approach increases the magnitude of premium sensitivity by about 44%, from -0.541 (SE = 0.118) to -0.777 (0.148). Premium interactions change in the IV model as well. The premium-by-FPL level coefficient increases from 0.000452 (0.000152) to 0.00769 (0.000177)—magnitudes are small here because FPL ranges from 138 to 400—and the premium-by-FPL over 200 coefficient increases from 0.204 (0.098) to 0.314 (0.117). Henceforth, we focus on the IV estimates.

**Appendix Table A3** reports first-stage estimates for each endogenous premium regressor. The instruments are strongly supported by F-tests of the three instruments in each regression, the lowest of which has an F-statistic of 5,440.67. Additionally, all instruments have coefficients that of sizeable magnitude and are significant in each regression.

## 4.6. Marginal Effects and Elasticities

We report calculated marginal effects and elasticities in **Table 3** for the IV specification overall and at standard FPL ranges in Marketplace analyses (138-150%, 151-200%, 201-250%, 251-300%, 301-400%). The marginal effect of a one dollar increase in the post-subsidy premium of the minimum premium silver plan on the probability of obtaining coverage is -0.40 (0.06) percentage points for the overall sample. Marginal effects decrease with income, from -0.67 (0.14) at 138-150% FPL with a sharp drop to -0.29 (0.04) at 200-250% FPL and -0.20 (0.03) at 301-400% FPL. This is consistent with a market bifurcated by the presence of cost-sharing reduction subsidies above and below 200% FPL (DeLeire et al. 2017).

While we report elasticities for consistency with the literature, we argue they are not particularly useful in our context. Specifically, elasticities become difficult to compare across income when prices for the same products exhibit exponential variation by income due to income-based subsidies. For example, the 301-400% FPL group has a mean post-subsidy premium that is several orders of magnitude higher for the highest income (235.65) than the 138-150% income group (1.65). This results in the higher income group having a premium elasticity of coverage of -1.28 (0.17) while the lowest group, counterintuitively, has an elasticity of -0.02 (0.00). This limitation does not apply to the overall elasticity, -0.75 (0.12), though we caution that this statistic describes the behavior of the market overall rather than individual enrollees.

Semi-elasticities eliminate this limitation of comparisons across income groups by examining the effect of one dollar change in post-subsidy premiums on coverage. The post-subsidy premium semi-elasticity of coverage for the 138-150% FPL group is -1.32 (0.27), indicating that a ten dollar increase in premiums for this group would decrease their enrollment by about 13%. Semi-elasticities are similar for the 151-200% FPL group (-1.21 (0.25)), decrease sharply for the 201-250% FPL group (-0.60 (0.08)), and continue to decrease at 301-400% FPL (-0.54 (0.07)). The overall semi-elasticity is -0.85 (0.14).

## 5. Optimal Subsidy Design

Having estimated demand for coverage, we next simulate how the elimination of ePTC subsidies will reduce coverage and how to optimally mitigate those losses with state supplemental subsidies. In doing so, we assume states' objectives are simply to maximize coverage subject to their budget constraints.

**5.1. Coverage Losses from Enhanced Premium Tax Credit Expiration and Costs to Restore**

We estimate coverage losses from ePTC expiration in two steps. First, we calculate the difference between the post-subsidy minimum premium silver plan with and without ePTCs. We then apply those differences to the estimated parameters from equation (1D) to predict 2024 coverage probabilities without ePTCs.

Our partial equilibrium results indicate that ePTC expiration would reduce subsidized MHBE enrollment from 136,308 to 80,065, a 41.3% reduction.[5] **Figure 1** displays these reductions across the income distribution. Coverage losses would be largest among enrollees with incomes at or below 200% FPL, among whom 58.3% (32,775) of the coverage losses will occur. A further 19,661 coverage losses would occur from 201-300% FPL (35.0%), with 3,807 losses (6.8%) from 301-400% FPL.

**5.2. Simulating Optimal Supplemental Premium Tax Credit Subsidy Allocation**

We now develop an algorithm that minimizes ePTC expiration coverage losses by iteratively assigning limited state supplemental subsidies to enrollees with the largest premium elasticities of coverage This objective is consistent with those of policymakers in states that have created supplemental subsidies, whom increasing premium affordability and coverage are key goals of the supplemental subsidies (Swindle and Giovanelli 2024).

The planner's goal, then, is to maximize retention of Marketplace enrollment among current enrollees. Each enrollee $i$ has a subsidy $s_i$ and a marginal effect of a one-dollar premium change on the probability of coverage, $dPr(Cvg_i)/dP_i$, obtained from the alpha coefficients in (1D). We define the planner's objective function as:

$$\max_{s} \sum_{i=1}^{N} s_i \frac{dPr(Cvg_i)}{dP_i} \qquad (3)$$

The planner is subject to a budget constraint set by state policymakers, and they will not make PTCs more generous than they were under the IRA. Formally, their constraints are:

---

[5] Practically, these enrollment reductions are likely to occur over a multi-year period rather than immediately. Enrollees may not immediately realize their premiums have increased (Drake et al. 2022), and some enrollment reductions will come from churn where current enrollees cycle off of the Marketplace and are not replaced by new ones that would enroll if ePTCs were available.

$$\sum_{i=1}^{N} s_i < B,$$

and

$$0 \leq s_i \leq s_i^{IRA} \ \forall i$$

Where $B$ is the supplemental subsidy budget and $s_i^{IRA}$ is an enrollee's IRA-level PTC. The algorithm naturally "fills" the subsidies of lower income, more premium sensitive enrollees before subsidizing higher income, less premium sensitive enrollees because marginal effects are constant for each enrollee based on their income.

We conduct our simulation in increments of $10 million annually, up to $150 million. These budgets are consistent with those MHBE has reviewed for reformulating its supplemental subsidies in 2026 (Maryland Health Benefit Exchange 2025). They will vary proportionately for other states based on the size and income distribution of their Marketplace enrollees.

### 5.3. Simulation Results

Our primary finding is that the planner can increase coverage by roughly 5,000 enrollees per year for $10 million in optimally allocated subsidies (i.e., to the most premium sensitive, lowest income enrollees first) until all enrollees with incomes below 200% FPL are fully subsidized. Up to 200% FPL, the mean enrollee receiving supplemental subsidies receives $160 per month. We report our simulations results in **Figure 2** and **Appendix Table A4**.

The cost-effectiveness of supplemental subsidies under 200% FPL—5,000 enrollees per $10 million budgeted—decreases beyond 200% FPL. In Maryland's case, this inflection point occurs slightly below $60 million dollars. It is important to note, though, that this $60 million is relative to MHBE's enrollment. Other states' thresholds will correspond to the dollar amount at which they have fully subsidized enrollees under 200% FPL, which will vary by states'

population sizes, income distributions, availability of employer-sponsored insurance, and Medicaid expansion decisions.[6]

The cost-effectiveness of supplemental subsidies dips by roughly 50% at 200% FPL, from 5,000 to 2,500 enrollees per $10 million, and gradually declines thereafter. **Figure 2** visualizes how enrollment increases with the budget, both overall (Panel A) and on the margin per $10 million budgeted (Panel B). Marginal enrollment declines to 2,500 enrollees on the margin between a $60 and $70 million budget (Panel B), when the income of the marginal enrollee with supplemental subsidies exceeds 200% FPL (Panel C). These declines in supplemental subsidies'' cost effectiveness continue as the budget and the marginal enrollee's income increase. Panels D and E show the cost effectiveness of subsidies at increasing overall and in $10 million increments for each budget; Panel F reports mean supplemental subsidy expenses per enrollee. Cost effectiveness declines with marginal enrollment with budgets starting at $70 million. Each additional $10 million in subsidies from $70 to $130 million results in 2,500 to 2,000 more enrollees and then declines rapidly as the marginal enrollees' income exceeds 350% FPL. Subsidy costs per enrollee also increase with budget size, from about $160 below with a $60 million budget covering <200% FPL enrollees, increasingly roughly linearly to $242.9 per enrollee at $150 million when all enrollees with incomes below 400% FPL receive supplemental subsidies.

**5.4. Discussion**

The primary finding of our simulations is that state supplemental subsidies to replace ePTCs are relatively cost-effective below 200% FPL. States with sufficient budgetary resources can retain coverage for approximately 5,000 lower income Marketplace enrollees that will otherwise lose coverage with ePTC expiration for about $10 million annually. Such returns are available until enrollees with incomes up to 200% FPL have ePTCs fully restored. At this point, about $60 million in Maryland's case, supplemental subsidies' cost effectiveness declines by roughly 50%, significantly diminishing the budgetary rationale for supplemental subsidies. States could potentially achieve more cost-effective returns on supplemental subsidies by fully subsidizing enrollees under 200% FPL, beyond ePTC levels and to the point where they have no premium, before subsidizing those with incomes above 200% FPL.[7] Whereas ePTCs already fully subsidized enrollees under 150% FPL, states could potentially expand the limit to 200% FPL.

---

[6] States that have not expanded Medicaid could subsidized enrollees from 100-138% FPL.

We note two other factors that states may wish to consider. First, maximizing access to care may imply a mixture of premium and cost-sharing subsidies. Lower income enrollees are highly sensitive to cost-sharing in deciding whether to pursue care, even when facing relatively small copays and deductibles under CSR 87% and 94% actuarial value silver plans (Lavetti et al. 2023). While premium subsidies may be sufficient for providing the financial protection health insurance provides, obtaining the health benefits of coverage may be contingent on eliminating cost-sharing for lower income enrollees. Several states have increased CSR subsidies through their Marketplaces or Basic Health Programs (Swindle and Giovanelli 2024; Zewde et al. 2024).

Second, there are conflicting reasons a state may wish to target older or younger enrollees. Targeting older enrollees could maximize the health returns of state subsidies. A large randomized study found large mortality decreases from Marketplace among older enrollees, but not younger ones (Goldin et al. 2020). In contrast, targeting younger enrollees may help to lower pre-subsidy premiums. If enrollees with incomes above 400% FPL lose access to PTCs, they will have to pay pre-subsidy premiums to obtain coverage. Bringing younger enrollees into the market could help to lower pre-subsidy premiums, which could be prohibitively costly to higher income enrollees otherwise. States should carefully consider the implications of such changes and their interactions with 1332 reinsurance waiver programs (Anderson et al. 2024).

---

[7] Many enrollees with incomes up to approximately 175% FPL could purchase the lowest silver premium plan for zero dollars after subsidies, though this still leaves those with incomes between 175-200% FPL that could be further subsidized.

### 5.5. Limitations of Simulation

We note several dimensions on which our simulations are partial equilibrium simulations. First, we do not model how insurers may change their plan offerings and prices in anticipation of and in response to changes in demand. This is a minor limitation insofar as the marginal enrollee is primarily concerned with post-subsidy premiums which are largely, though not entirely, determined by subsidies and not insurer plan offerings. Second, we do not model how additional subsidies affect enrollees' plan choices and, in turn, how much of their subsidies they use. Enrollees selecting low- or zero-dollar coverage may have residual, unused subsidies. In this sense, our subsidy cost estimates are an upper bound. Third, we do not consider how subsidy changes may affect duration of enrollment. Lastly, we do not model the effects of other federal policy changes that will likely decrease enrollment and thereby supplemental subsidy costs, notably income verification requirements to be implemented in 2027 (Kaiser Family Foundation 2025).

## 6. Conclusions

We provide new, post-ARPA/IRA estimates of demand for Marketplace coverage using administrative data from Maryland's Health Insurance Marketplace. We then simulate supplemental subsidies effects on enrollment and state budgetary costs under ePTC expiration. States could feasibly implement subsidies to mitigate ePTC coverage losses among Marketplace enrollees with incomes at or below 200% FPL, where it costs the state roughly $10 million per

year to increase coverage by 5,000 enrollees. The cost effectiveness of the subsidies drops by roughly half for higher income enrollees. Given that lower income enrollees' higher premium sensitivities make them more vulnerable to ePTC cuts and that mitigating their coverage losses can be partially accomplished for what states are currently spending on supplemental Marketplace subsidies, states interested in preserving this decade's Marketplace coverage gains should consider creating supplemental subsidies for lower income Marketplace enrollees.

**Declaration of generative AI and AI-assisted technologies in the manuscript preparation process**

During the preparation of this work the authors used ChatGPT 5.0 to assist with simulation coding. After using this tool/service, the authors reviewed and edited the content as needed and take full responsibility for the content of the published article.

**Figure 1** Projected Coverage Losses from ePTC Expiration

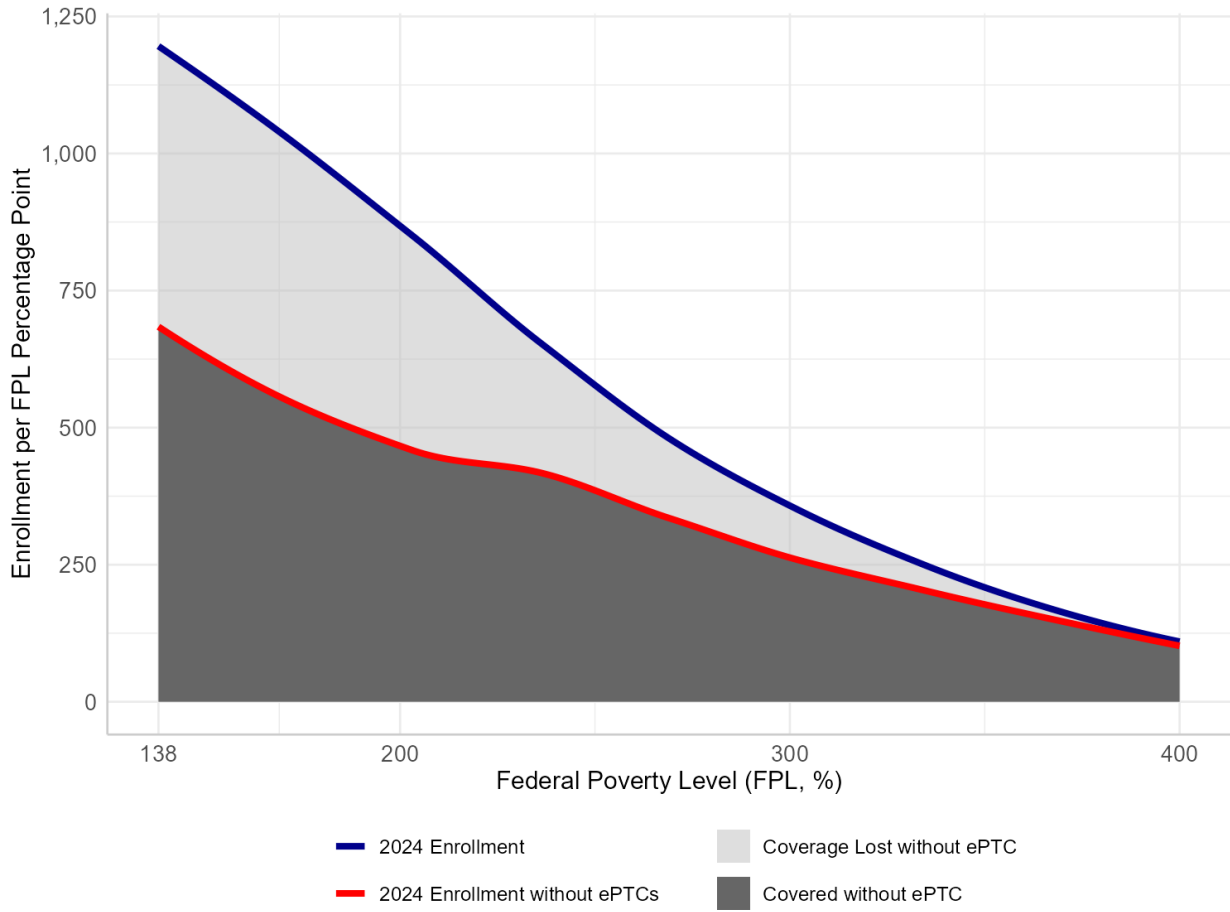

*Notes.* 2024 enrollment includes all subsidized MHBE enrollees with incomes from 138-400% FPL without missing data (i.e., our analytic sample for 2024). We calculate 2024 enrollment without ePTCs by: (1) calculating changes in premium subsidies and post-subsidy premiums using pre-ARPA 2021 expected contribution percentages prior to ePTC implementation; and (2) multiplying those changes by the marginal effects for premiums show in our primary IV specification shown in Table 2. Table 3 reports marginal effects of premium changes at key points on the income distribution.

**Figure 2** Coverage Gains and Costs of Supplemental Marketplace Premium Subsidies by Income

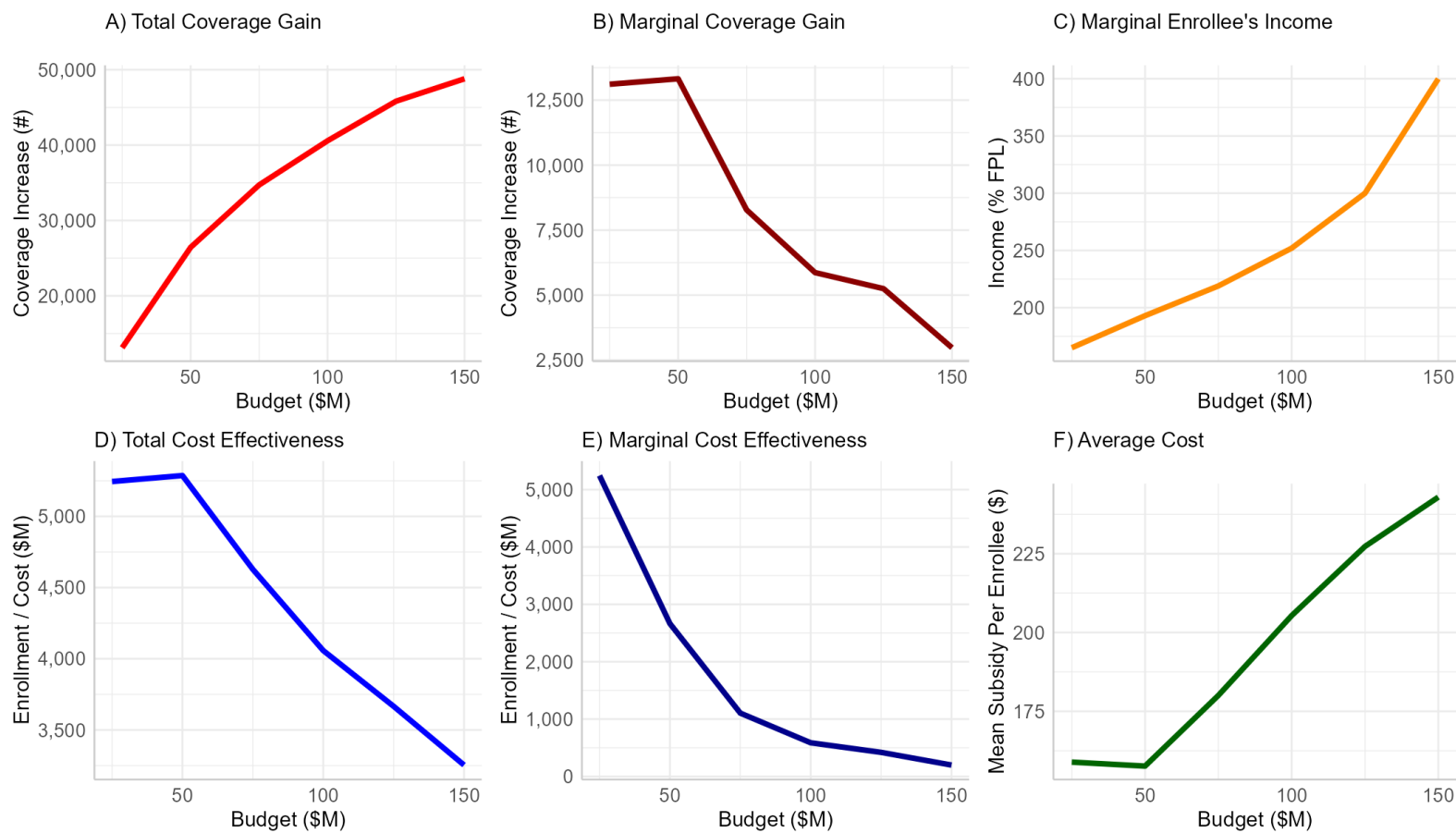

*Notes.* Each panel shows the results of simulations conducted in intervals of $5 million for budgets ranging from $10 million to $150 million annually. Panel A shows the number of enrollees that retain coverage due to supplemental state subsidies. Panel B displays the income of the marginal enrollee that maintains coverage that receives state supplemental subsidies. Panel C visualizes the average annual cost per enrollee of state supplemental subsidies. Panel D graphs the cost to the state in supplemental subsidies of increasing the marginal enrollee's probability of coverage by one percentage point. See Section 5.2 for simulation details. See Appendix Table A4 for tabular results.

**Table 1** Choice and Demographic Distributions of Maryland Health Benefit Exchange Enrollees, 2022-2024

|  | Year, Mean (SD) | | | |
|---|---|---|---|---|
| Characteristic (SD) | 2022 | 2023 | 2024 | All |
| Enrollment (1,000's) | | | | |
|   Individuals | 110.9 | 122.2 | 143.1 | 216.1 |
|   Households | 86.3 | 94.5 | 111.9 | 169.1 |
| Demographics | | | | |
|   Income (% FPL) | 230.2 (67.8) | 230.2 (65.9) | 223.6 (64.7) | 227.7 (66.1) |
|   Age (Years) | 44.1 (13.6) | 43.3 (13.8) | 42.3 (14) | 43.1 (13.9) |
|   Female (%) | 55.8 | 57.2 | 57.9 | 57.0 |
| Subsidies and Plan Selection | | | | |
|   Premium ($) | | | | |
|     Pre-PTC | 703.4 (415.6) | 731 (436.5) | 728.1 (449.2) | 721.8 (435.6) |
|     Post-PTC | 162.0 (177.2) | 184.9 (198.4) | 174.4 (205.4) | 174.1 (195.4) |
|   Federal PTC ($) | | | | |
|     Awarded | 554.4 (369.4) | 556.9 (376.5) | 565 (389.1) | 559.3 (379.3) |
|     Applied | 541.6 (362.5) | 546.6 (371.5) | 554.2 (383.5) | 548.0 (373.6) |
|   Maryland YAS PTC ($) | | | | |
|     Awarded | 52.6 (30.8) | 46.9 (37.6) | 51.0 (43.5) | 50.1 (39.0) |
|     Applied | 47.9 (31.1) | 44.7 (37.2) | 47.9 (42.1) | 46.9 (38.2) |
|   Metal Level (%) | | | | |
|     Bronze | 21.4 | 22.5 | 22.2 | 22.1 |
|     Silver | 27.1 | 27.0 | 26.5 | 26.8 |
|     Gold | 49.3 | 48.6 | 49.8 | 49.3 |
|     Platinum | 2.2 | 2.0 | 1.5 | 1.9 |

*Notes*. Sample includes MHBE enrollees receiving premium tax credits from 2022 through 2024. In 2022, the Maryland Young Adult subsidy program reduced expected contribution percentages by 2.5 percentage points for adults aged 18 to 30, phasing out by 0.5 points per year from ages 31 to 35. In 2023, the program was fully expanded to age 34, phasing out at age 38. The program is in the process of being redesigned for 2026.

**Table 2** Ordinary Least Squares and Two-Stage Least Squares Baseline Regression Results

|  | Specification | |
| --- | --- | --- |
| Covariate | OLS | 2SLS (IV) |
| Premium Sensitivity: Instrumented in IV Models | | |
|   Premium | -0.541 (0.118) | -0.777 (0.148) |
|   Interaction: Premium, FPL (Continuous) | 0.000452 (0.000152) | 0.00769 (0.000177) |
|   Interaction: Premium, FPL >200% (0/1) | 0.204 (0.098) | 0.314 (0.117) |
| Demographics | | |
|   Female | 16.050 (1.246) | 15.922 (1.248) |
|   Age | 1.028 (0.077) | 1.105 (0.089) |
|   FPL | 0.125 (0.029) | 0.138 (0.033) |
| Rating Area-by-Year Fixed Effects | Yes | Yes |
| N | 387,076 | 387,076 |

*Notes*. FPL, a continuous measure of income as a percentage of the federal poverty level, is z-score normalized according to its weighted mean and SD. This is also the case for its squared and cube, as well as premium interactions. Standard errors are clustered by health insurance units, sub-household units that jointly purchase health insurance. See Appendix Table A3 for first stage IV results.

**Table 3** Marginal Effects, Semi-Elasticities, and Elasticities by Income

| Income Group (% FPL) | Mean Annual Enrollment (Count) | Weighted Means[1] | | Coverage Responses[2] | | |
|---|---|---|---|---|---|---|
| | | Monthly Premium ($) | Enrollment Rate (%) | Marginal Effect[3] (pp) | Semi-Elasticity[4] | Elasticity[5] |
| 138-400 (All) | 129,025 | 87.52 | 46.97 | -0.40 (0.06) | -0.85 (0.14) | -0.75 (0.12) |
| 138-150 | 11,632 | 1.65 | 50.35 | -0.67 (0.14) | -1.32 (0.27) | -0.02 (0.00) |
| 151-200 | 43,412 | 11.03 | 53.29 | -0.64 (0.13) | -1.21 (0.25) | -0.13 (0.03) |
| 201-250 | 31,671 | 51.44 | 48.07 | -0.29 (0.04) | -0.60 (0.08) | -0.31 (0.04) |
| 251-300 | 20,779 | 129.32 | 45.98 | -0.25 (0.03) | -0.55 (0.07) | -0.71 (0.10) |
| 301-400 | 21,531 | 235.65 | 36.42 | -0.20 (0.03) | -0.54 (0.07) | -1.28 (0.17) |

[1] Uninsured Marketplace subsidy-eligible persons weighted with American Community Survey sample weights.
[2] Calculated using weighted means and IV estimates from Table 2 using the delta method for standard errors.
[3] The percentage change in the probability of coverage resulting from a one dollar increase in the post-subsidy monthly premium of the lowest premium silver plan.
[4] The percent change in the probability of coverage resulting from a one dollar increase in the post-subsidy monthly premium of the lowest premium silver plan.
[5] The percent change in the probability of coverage resulting from a one percent increase in the post-subsidy monthly premium of the lowest premium silver plan. The relationship with pure elasticities and income is inconsistent because the weighted means of monthly premiums exhibit large variation across the income distribution.

**Appendix Figure A1** Expected Contribution Percentages and Monthly Contributions by Income, ACA vs. IRA

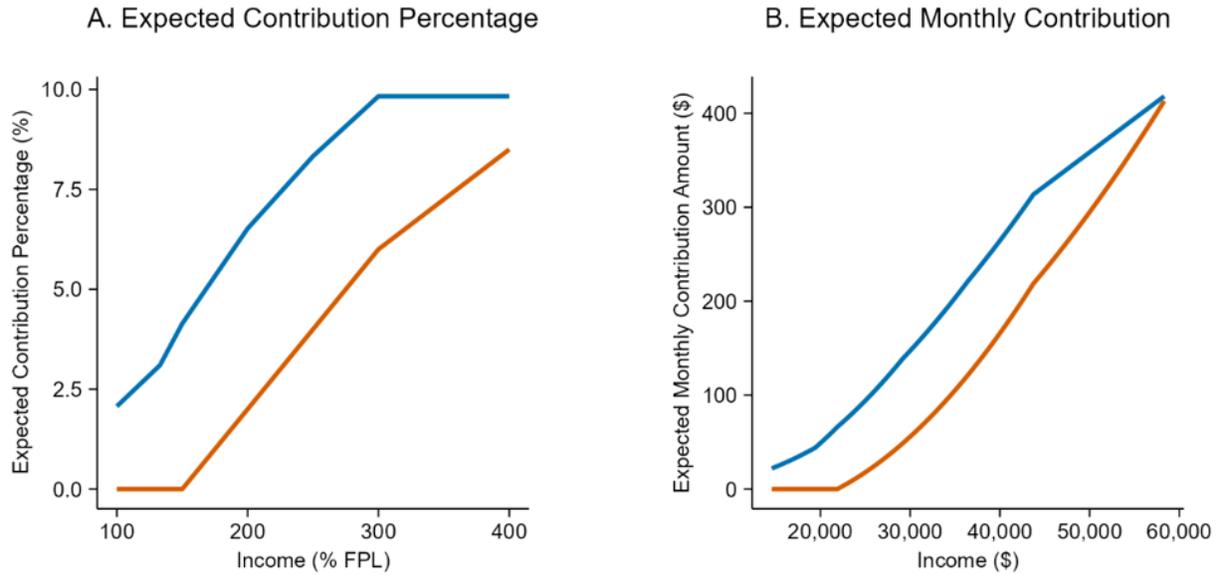

*Notes*. The ACA subsidy regime is the original subsidy regime created by the ACA in 2014 to which states will revert in 2026. The IRA subsidy regime increased subsidy generosity from 2022-2025. Subsidies for enrollees with incomes >400% FPL are only available under the IRA subsidy regime and are not shown here.

**Appendix Table A1**. State Supplemental Premium Subsidy Approaches

| State | Years | Description |
| --- | --- | --- |
| California | 2020-21 | ECP capped up to 600% FPL |
| Connecticut | ≥2021 | Reduce premiums to zero ≤175% FPL |
| Massachusetts | ≥2007 | ECP reduction based on income up to 300% FPL |
| Minnesota | 2017 | Reduce premium by 25% if not subsidized |
| Maryland* | 2022-25 | Reduce ECP by 2.5pp ages 18-30, phaseout to age 35 |
| New Jersey | ≥2021 | ECP reduction based on income |
| New Mexico | ≥2023 | ECP reduction based on income |
| Washington | ≥2023 | Fixed dollar reduction based on income |
| Vermont | ≥2014 | Reduce ECP by 1.5pp ≤300% FPL |

*Notes.* Information obtained from State Health and Value Strategies and the Commonwealth Fund (Levitis and Pandit 2021; Swindle and Giovanelli 2024). California has state subsidies that cover the one dollar per member-month cost of mandatory abortion benefits; Maryland allowed it Young Adult Subsidy to cover non-Essential Health Benefits beginning in 2023. Minnesota, New York, and Oregon have Basic Health Programs that have reduced premiums for enrollees with incomes below 200% FPL. Colorado provides subsidies for some immigrants otherwise ineligible for premium subsidies. The District of Columbia fully subsidies post-subsidy premiums for childcare facility enrollees. * Age limits were increased to 18-32 with phase out in 37 beginning in 2023. Maryland will redesign its subsidies for the 2026 plan year.

**Appendix Table A2** Demographic Comparison of MHBE Enrollees and Potential Enrollees

| Characteristic | MHBE Enrollees | Uninsured Potential Enrollees |
| --- | --- | --- |
| Population (1,000s) | | |
|   Overall | 384.1 | 433.6 |
|   2022 | 113.2 | 138.1 |
|   2023 | 124.6 | 147.7 |
|   2024 | 146.2 | 147.7 |
| Demographics | | |
|   Income (FPL %) | 227.7 (66.1) | 244.6 (72.1) |
|   Age (Years) | 43.1 (13.9) | 38.7 (11.6) |
|   Female (%) | 57.0 | 38.3 |
| Subsidies and Plan Availability | | |
|   Federal PTC awarded | 351.9 (198.1) | 254.2 (164.5) |
|   Post-PTC Min. Silver Premium | 92.4 (93.8) | 117.5 (104.7) |

*Notes*. Samples include (1) MHBE enrollees receiving premium tax credits from 2022 through 2024, and (2) potential, subsidy-eligible MHBE enrollees identified in the American Community Survey (ACS) from 2022 through 2024. ACS data are currently only available through 2023, so we imputed 2024 data as 2023 data.

**Appendix Table A3** First Stage Estimates

| Endogenous Regressor | Covariate (Instrument) | Estimate (SE) | F-Statistic |
|---|---|---|---|
| ARPA Premium | ACA Premium | -0.351 (0.012) | 5440.67 |
|  | ACA Premium, FPL | 0.003 (0.000) |  |
|  | ACA Premium, FPL >200% | -0.008 (0.004) |  |
| ARPA Premium, FPL (Continuous) | ACA Premium | -323.293 (3.473) | 10403.32 |
|  | ACA Premium, FPL | 1.711 (0.011) |  |
|  | ACA Premium, FPL >200% | -26.474 (1.233) |  |
| ARPA Premium, FPL > 200% (0/1) | ACA Premium | -0.675 (0.012) | 6938.55 |
|  | ACA Premium, FPL | 0.003 (0.000) |  |
|  | ACA Premium, FPL >200% | 0.218 (0.005) |  |

*Notes*. These estimates are first stage estimates for IV results shown in Appendix Table A2. FPL, a continuous measure of income as a percentage of the federal poverty level, is z-score normalized according to its weighted mean and SD. This is also the case for its squared and cube, as well as premium interactions. Standard errors are clustered by health insurance units, sub-household units that jointly purchase health insurance.

**Appendix Table A4** Simulation Coverage Gains and Annual Costs by Budget, 10 to 150 Million Dollars Per Year

| Budget | Coverage Gain (1,000's Enrollees) | | Marginal Enrollee's Income (% FPL) | Cost Effectiveness (Enrollment / $M) | | Mean Subsidy Per Enrollee ($) |
| --- | --- | --- | --- | --- | --- | --- |
| | Total | Marginal | | Total | Marginal | |
| 10 | 5.1 | 5.1 | 150 | 5,100 | 5,100 | 162.4 |
| 20 | 10.4 | 5.3 | 160 | 5,200 | 5,300 | 160.3 |
| 30 | 15.8 | 5.4 | 171 | 5,267 | 5,400 | 158.2 |
| 40 | 21.2 | 5.4 | 181 | 5,300 | 5,400 | 157.6 |
| 50 | 26.4 | 5.2 | 193 | 5,280 | 5,200 | 157.6 |
| 60 | 31 | 4.6 | 203 | 5,167 | 4,600 | 161.5 |
| 70 | 33.5 | 2.5 | 213 | 4,786 | 2,500 | 174.3 |
| 80 | 35.9 | 2.4 | 226 | 4,488 | 2,400 | 185.6 |
| 90 | 38.3 | 2.4 | 240 | 4,256 | 2,400 | 195.9 |
| 100 | 40.6 | 2.3 | 252 | 4,060 | 2,300 | 205.4 |
| 110 | 42.8 | 2.2 | 267 | 3,891 | 2,200 | 214.3 |
| 120 | 44.8 | 2 | 286 | 3,733 | 2,000 | 223 |
| 130 | 46.8 | 2 | 312 | 3,600 | 2,000 | 231.6 |
| 140 | 48.5 | 1.7 | 356 | 3,464 | 1,700 | 240.6 |
| 150 | 48.8 | 0.3 | 400 | 3,253 | 300 | 242.9 |

*Notes.* See Section 5.2 for simulation details. See Figure 2 for illustrated results. * The amount the state must spend in supplemental subsidies to increase the probability that the marginal enrollee covered by subsidies will increase their probability of enrolling in coverage by one percentage point. This quantity increases because premium sensitivity declines with income.